\documentclass[12pt]{article}
\usepackage{amsfonts}
\usepackage{amssymb}
\usepackage{subfigure}
\usepackage{graphicx}
\usepackage{verbatim}

\def\hybrid{\topmargin -30pt    \oddsidemargin 0pt 
        \headheight 0pt \headsep 0pt
        \textwidth 6.25in       
        \textheight 9.5in       
        \marginparwidth .875in
        \parskip 5pt plus 1pt   \jot = 1.5ex}

\hybrid

\def\baselinestretch{1.2}

\catcode`\@=11

\def\marginnote#1{}
\newcount\hour
\newcount\minute
\newtoks\amorpm
\hour=\time\divide\hour by60
\minute=\time{\multiply\hour by60 \global\advance\minute by-\hour}
\edef\standardtime{{\ifnum\hour<12 \global\amorpm={am}%
        \else\global\amorpm={pm}\advance\hour by-12 \fi
        \ifnum\hour=0 \hour=12 \fi
        \number\hour:\ifnum\minute<10 0\fi\number\minute\the\amorpm}}
\edef\militarytime{\number\hour:\ifnum\minute<10 0\fi\number\minute}

\def\draftlabel#1{{\@bsphack\if@filesw {\let\thepage\relax
   \xdef\@gtempa{\write\@auxout{\string
      \newlabel{#1}{{\@currentlabel}{\thepage}}}}}\@gtempa
   \if@nobreak \ifvmode\nobreak\fi\fi\fi\@esphack}
        \gdef\@eqnlabel{#1}}
\def\@eqnlabel{}
\def\@vacuum{}
\def\draftmarginnote#1{\marginpar{\raggedright\scriptsize\tt#1}}

\def\draft{\oddsidemargin -.5truein
        \def\@oddfoot{\sl preliminary draft \hfil
        \rm\thepage\hfil\sl\today\quad\militarytime}
        \let\@evenfoot\@oddfoot \overfullrule 3pt
        \let\label=\draftlabel
        \let\marginnote=\draftmarginnote
   \def\@eqnnum{(\theequation)\rlap{\kern\marginparsep\tt\@eqnlabel}%
\global\let\@eqnlabel\@vacuum}  }

\def\draft2{
        \def\@oddfoot{\sl preliminary draft \hfil
        \rm\thepage\hfil\sl\today\quad\militarytime}
        \let\@evenfoot\@oddfoot \overfullrule 3pt
        \let\label=\draftlabel
        \let\marginnote=\draftmarginnote
   \def\@eqnnum{(\theequation)\rlap{\kern\marginparsep\tt\@eqnlabel}%
\global\let\@eqnlabel\@vacuum}  }

\def\preprint{\twocolumn\sloppy\flushbottom\parindent 2em
        \leftmargini 2em\leftmarginv .5em\leftmarginvi .5em
        \oddsidemargin -.5in    \evensidemargin -.5in
        \columnsep .4in \footheight 0pt
        \textwidth 10.in        \topmargin  -.4in
        \headheight 12pt \topskip .4in
        \textheight 6.9in \footskip 0pt
        \def\@oddhead{\thepage\hfil\addtocounter{page}{1}\thepage}
        \let\@evenhead\@oddhead \def\@oddfoot{} \def\@evenfoot{} }

\def\numberbysection{\@addtoreset{equation}{section}
        \def\theequation{\thesection.\arabic{equation}}}

\def\underline#1{\relax\ifmmode\@@underline#1\else
        $\@@underline{\hbox{#1}}$\relax\fi}

\def\titlepage{\@restonecolfalse\if@twocolumn\@restonecoltrue\onecolumn
     \else \newpage \fi \thispagestyle{empty}\c@page\z@
        \def\thefootnote{\fnsymbol{footnote}} }

\def\endtitlepage{\if@restonecol\twocolumn \else \newpage \fi
        \def\thefootnote{\arabic{footnote}}
        \setcounter{footnote}{0}}  

\catcode`@=12
\relax

\def\figcap{\section*{Figure Captions\markboth
        {FIGURECAPTIONS}{FIGURECAPTIONS}}\list
        {Figure \arabic{enumi}:\hfill}{\settowidth\labelwidth{Figure
999:}
        \leftmargin\labelwidth
        \advance\leftmargin\labelsep\usecounter{enumi}}}
 \relax
\def\tablecap{\section*{Table Captions\markboth
        {TABLECAPTIONS}{TABLECAPTIONS}}\list
        {Table \arabic{enumi}:\hfill}{\settowidth\labelwidth{Table
999:}
        \leftmargin\labelwidth
        \advance\leftmargin\labelsep\usecounter{enumi}}}
 \relax
\def\reflist{\section*{References\markboth
        {REFLIST}{REFLIST}}\list
        {[\arabic{enumi}]\hfill}{\settowidth\labelwidth{[999]}
        \leftmargin\labelwidth
        \advance\leftmargin\labelsep\usecounter{enumi}}}
 \relax
\makeatletter
\newcounter{pubctr}
\def\publist{\@ifnextchar[{\@publist}{\@@publist}}
\def\@publist[#1]{\list
        {[\arabic{pubctr}]\hfill}{\settowidth\labelwidth{[999]}
        \leftmargin\labelwidth
        \advance\leftmargin\labelsep
        \@nmbrlisttrue\def\@listctr{pubctr}
        \setcounter{pubctr}{#1}\addtocounter{pubctr}{-1}}}
\def\@@publist{\list
        {[\arabic{pubctr}]\hfill}{\settowidth\labelwidth{[999]}
        \leftmargin\labelwidth
        \advance\leftmargin\labelsep
        \@nmbrlisttrue\def\@listctr{pubctr}}}
 \relax
\makeatother

\def\be{\begin{equation}}
\def\ee{\end{equation}}
\def\ba{\begin{eqnarray}}
\def\ea{\end{eqnarray}}

\def\del{\partial}

\def\r{\rho}
\def\a{\alpha}

\def\b{\beta}

\def\g{\gamma}
\def\G{\Gamma}
\def\d{\delta}
\def\D{\Delta}
\def\e{\epsilon}

\def\m{\mu}
\def\n{\nu}

\def\l{\lambda}

\def\s{\sigma}
\def\S{\Sigma}

\def\no{\noindent}

\def\qq{\qquad}

\def\IR{\relax{\rm I\kern-.18em R}}

\def\tT{{\tilde T}}
\def\tg{{\tilde g}}

\def\tL{{\tilde L}}
\def\PL{Poisson--Lie T-duality}

\def\inv{^{\raise.0ex\hbox{${\scriptscriptstyle -}$}\kern-.05em 1}}

\def \ha {{1\over 2}}

\def \ov {\over}

\begin{document}


\renewcommand{\theequation}{\thesection.\arabic{equation}}
\csname @addtoreset\endcsname{equation}{section}

\newcommand{\eqn}[1]{(\ref{#1})}
\begin{titlepage}
\begin{center}

\phantom{xx}
\vskip 0.5in

{\Large \bf Quantum equivalence in Poisson-Lie T-duality}

\vskip 0.5in

{\bf Konstadinos Sfetsos}\phantom{x} and\phantom{x}
{\bf Konstadinos Siampos}
\vskip 0.1in

Department of Engineering Sciences, University of Patras,\\
26110 Patras, Greece\\

\vskip .1in

\vskip .15in

{\footnotesize {\tt sfetsos@upatras.gr},
\ \ {\tt ksiampos@upatras.gr}}\\

\end{center}

\vskip .4in

\centerline{\bf Abstract}

\no
We prove that, general $\s$-models related by Poisson--Lie T-duality are quantum
equivalent under one-loop renormalization group flow.
We reveal general properties of the flows, we study
the associated generalized coset models and provide explicit examples.

\no


\end{titlepage}
\vfill
\eject


\tableofcontents

\def\baselinestretch{1.2}
\baselineskip 20 pt
\no

\section{Introduction}

A generalization of the well known Abelian \cite{BUSCHER} and non-Abelian \cite{nonabel}
T-dualities is the so-called \PL\ \cite{KliSevI}.
Its most notable feature is that it does not
rely on the existence of isometries but rather on a
rigid group-theoretical structure \cite{KliSevI}. Nevertheless, it shares some common
features with ordinary T-duality. For instance, it can
be explicitly formulated as a canonical transformation between phase-space
variables \cite{PLsfe1,PLsfe2}, similarly to ordinary T-duality
\cite{zacloz, AALcan}, a property that seems to be very important
for our considerations.\footnote{Related developments include
works on open string boundary conditions (see, for instance,
\cite{KlimcikDbra, Albertsson}).}

\no
An important question that
was addressed successfully in a particular example in \cite{PLsfe3}
is whether classically equivalent, via canonical transformations, models retain
their equivalence beyond the classical level in the following sense.
As two-dimensional field theories, these $\s$-models are renormalizable if the corresponding
counter-terms, at a given order in a loop expansion,
can be absorbed into a renormalization of the various coupling constants appearing in the model up
to field redefinitions or, equivalently, diffeomorphisms in the target space.
These give rise to beta-function renormalization group equations which
generally form a non-linear coupled system of first order differential equations.
The classical equivalence of the two models can be promoted order by order in perturbation theory
into the quantum level if the two different renormalization group flow systems of
equations are in fact equivalent. In turn, this strongly hints towards their
equivalence beyond the classical level. The existence of a canonical transformation relating two
different $\s$-models seems to be
necessary for their equivalence at the quantum
level.\footnote{As a counterexample note the case of the Principal Chiral model and the
Pseudodual Chiral model \cite{pscm} whose classical solutions are in one-to-one
correspondence but which are not canonically equivalent \cite{zacloz}.
It is well known
that their quantum behaviors are drastically different \cite{Nappi}.}
A general proof for all \PL\ related models requires
the generalized Ricci tensor. The latter was computed in \cite{ValCli} where it was
shown that \PL\ related models are renormalizable in the above described sense.

\no
The main objective of the present work is to show the equivalence of the renormalization group flow
for general $\s$-models related by \PL.
We study properties of the flows and prove several general statements
regarding mainly the possible truncations of the parameter
space in a way consistent with the renormalization group equations.
We also
investigate the generalized coset models introduced in \cite{PLsfe4}
at the purely classical level.
Finally, we present some explicit examples
based on three-dimensional algebras.

\section{Brief review of \PL\ T-duality}

In the section we review the most relevant aspects of \PL\ for our purposes by following the
conventions of \cite{PLsfe1,PLsfe2,PLsfe3,PLsfe4}.
In the absence of spectator fields, the dual two-dimensional $\s$-model actions
are \cite{KliSevI}
\be
S= {1\ov 2 \l} \int E_{ab} L^a_\m L^b_\n \del_+ X^\m \del_- X^\n\ ,\qq E= (M - \Pi)^{-1}\ ,
\label{action1}
\ee
and
\be
\tilde S= {1\ov 2 \l} \int \tilde E^{ab} \tL_{a\m} \tL_{\b\n} \del_+
 \tilde X^\m \del_- \tilde X^\n\ ,\qq \tilde E=(M^{-1} - \tilde \Pi)^{-1}\ .
\label{action2}
\ee
The field variables $X^\m$ and $\tilde X^\m$, with $\m=1,2,\dots  ,d_G$ parametrize elements
$g$ and $\tilde g$ of two groups $G$ and $\tilde G$, respectively, of equal dimension.
We introduce representation matrices, $\{T_a\}$ and $\{\tilde T^a\}$
with $a=1,2,\dots, d_G$,
of the associated Lie algebras which form a pair
of maximally isotropic subalgebras into which the Lie algebra of a
group, known as the Drinfeld double, can be decomposed.\footnote{
A generalization of \PL\ named Poisson-Lie T-plurality \cite{VonUnge}-\cite{Hlavaty}
is based on the non-uniqueness of the decomposition of the Drinfeld double.
}
The commutation relations are
\ba
&& [T_a,T_b]  =  i f_{ab}{}^c T_c \ ,
\nonumber\\
&& [\tilde T^a, \tilde T^b]  =  i \tilde f^{ab}{}_c \tilde T^c \ ,
\\
&& [T_a,\tilde T^b ] = i \tilde f^{bc}{}_a T_c - i f_{ac}{}^b \tilde T^c
\nonumber
\ea
and imply the Jacobi identities (in our conventions $(ab)=ab+ba$ and $[ab]=ab-ba$)
\ba
&& f_{ab}{}^d  f_{dc}{}^e + f_{ca}{}^d  f_{db}{}^e + f_{bc}{}^d  f_{da}{}^e = 0 \ ,
\nonumber\\
&& \tilde f^{ab}{}_d  \tilde f^{dc}{}_e + \tilde f^{ca}{}_d  \tilde f^{db}{}_e
+ \tilde f^{bc}{}_d  \tilde f^{da}{}_e = 0 \ ,
\label{ftilf1}\\
&& f_{ab}{}^d \tilde f^{ce}{}_d + f_{d[a}{}^c \tilde f^{de}{}_{b]}
- f_{d[a}{}^e \tilde f^{dc}{}_{b]} = 0 \ .
\nonumber
\ea
Introducing a bilinear invariant
$\langle{\cdot|\cdot \rangle}$ with the various representation matrices obeying
\be
\langle{T_a|T_b\rangle}= \langle{\tilde T^a|\tilde T^b \rangle}= 0\ ,
\qq \langle{T_a|\tilde T^b \rangle} = \d_a{}^b \ ,
\label{bili}
\ee
we define the components of the Maurer--Cartan forms appearing in the above actions as
\be
L^a_\m =-i  \langle g\inv \del_\m g| \tT^a \rangle\ ,
\qq \tL_{a\m} =  -i\langle \tg\inv \del_\m \tg| T_a \rangle\ .
\ee

\no
The overall coupling constant is $\l$ in \eqn{action1} and \eqn{action2} is assumed to be positive and
the square matrix $M$ is constant and has dimension $d_G$.
The matrices
$\Pi$ and $\tilde \Pi$ depend on the variables $X^\m$ and $\tilde
X^\m$ via the corresponding group elements $g$ and $\tilde g$, respectively.
They are defined as
\be
\Pi^{ab} = b^{ca} a_c{}^b \ , \qq
\tilde \Pi_{ab} = \tilde b_{ca} \tilde a^c{}_b \ ,
\label{pipi}
\ee
where the matrices $a(g)$, $b(g)$ are constructed using
\be
g\inv T_a g = a_a{}^b T_b\ ,\qq g\inv \tilde T^a g =
b^{ab} T_b +  (a\inv)_b{}^a \tilde T^b\ .
\label{abpi}
\ee
Consistency restricts them to obey
\be
a(g\inv) = a\inv(g)\ ,\qq b^T(g)= b(g\inv)\ ,\qq
\Pi^T(g) = - \Pi(g) \ ,
\label{conss}
\ee
We also define the dual tilded symbols with similar properties.

\section{Renormalization group and Poisson--Lie T-duality}

We begin this section with a short review of the renormalization group in
two-dimensional field theories with curved target spaces.
The $\s$-models \eqn{action1} and \eqn{action2} are of the form
\be
S={1\ov 2\l} \int Q^+_{\m\n} \del_+ X^\m \del_- X^\n\ ,
\qq Q^+_{\m\n}\equiv G_{\m\n} + B_{\m\n}\ .
\label{fjll}
\ee
As a two-dimensional field theory for the fields $X^\m$,
this will be renormalizable if the corresponding counter-terms, at
a given order in a loop expansion,
can be absorbed into a renormalization of the
coupling constant $\l$ and (or) of
some parameters labeled collectively by $a^i$, $i=1,2,\dots$. In our case these are the
entries of the matrix $M^{ab}$ and the overall coupling constant $\l$.
In doing so, we may allow for general field redefinitions of the $X^\m$'s,
which are coordinate reparametrizations in the target space.
This definition of renormalizability of $\s$-models is quite strict and
similar to that for ordinary field theories. An extension of this is to
allow for the manifold to vary with the mass scale and the renormalization group
to act in the infinite-dimensional space of all metrics and
torsions \cite{Frialv}, but this will not be needed for our purposes.

\no
The one-loop beta-functions are expressed as
\ba
 {d\l\ov dt} = {\l^2\ov \pi} J_1\ ,
\qq
{da^i\ov dt}= {\l \ov \pi} a^i_1\ ,
\ea
where $t=\ln \m$, with $\m$ being the mass energy scale and $J_1,a^i_1$ and $X_1^\m$
are chosen so that
\be
\ha R^-_{\m\n} =
-J_1 Q^+_{\m\n} + \del_{a^i} Q^+_{\m\n} a^i_1 + \del_\l Q^+_{\m\n} X_1^\l
+  Q^+_{\l\n} \del_\m  X_1^\l  +  Q^+_{\m\l} \del_\n X_1^\l\ .
\label{1loop}
\ee
Here $R^-_{\m\n}$ are the components of
the generalized Ricci tensor defined with a connection
that includes the torsion, i.e. $\G^{\m}_{\n\r} - \ha H^\m{}_{\n\r}$.
The counter-terms were computed in \cite{vanderven,Frialv,Osborn}.

\subsection{One loop renormalization group for Poisson--Lie T-duals}

For the Poisson--Lie T-duality related actions \eqn{action1} and \eqn{action2}
it was first shown that it is possible to satisfy the system \eqn{1loop} in the
case of a six-dimensional Drinfeld double
in \cite{PLsfe3,PLsfe4}.\footnote{
For Abelian
and non-Abelian dualities, similar investigation were performed
for a one-parameter family deformations
of the Principal Chiral model for $SU(2)$,
and its non-Abelian dual \cite{zacloz,ouggroi}, in \cite{ouggroi}.

}
This made plausible that
it could be true for at least a large class of such doubles.
A general proof requires the computation of
the generalized Ricci tensor for each $\s$-model separately.
Using the underlying Poisson--Lie structure and in particular
some useful identities derived in \cite{PLsfe2}
this was done in \cite{ValCli},\footnote{To compare our conventions with those in
\cite{ValCli}
one should perform the following replacements in that paper.
Namely, we send $g\to g^{-1}$, $\Pi\to -\Pi$ and $\tilde \Pi\to -\tilde \Pi$ .}
where it was also shown that the models are one-loop
renormalizable in the sense explained above.

\no
We define for convenience the following quantities
\be
A^{ab}{}_{c} = \tilde f^{ab}{}_c - f_{cd}{}^a M^{db}\ ,\qq
B^{ab}{}_{c} = \tilde f^{ab}{}_c + M^{ad}f_{dc}{}^b \ ,
\label{fhh11}
\ee
as well as their duals
\be
\tilde A_{ab}{}^{c} =  f_{ab}{}^c - \tilde f^{cd}{}_a (M^{-1})_{db}\ ,\qq
\tilde B_{ab}{}^{c} = f_{ab}{}^c + (M^{-1})_{ad}\tilde f^{dc}{}_b \ ,
\ee
Using these we construct also
\ba
L^{ab}{}_c & = &  \ha (M_s^{-1})_{cd}\left( B^{ab}{}_e M^{ed} + A^{db}{}_e M^{ae}- A^{ad}{}_eM^{eb}
  \right) \ ,
\nonumber\\
R^{ab}{}_c & = & \ha (M_s^{-1})_{cd}\left( A^{ab}{}_e M^{de} + B^{ad}{}_eM^{eb} - B^{db}{}_e M^{ae}
\right) \
\label{kkll1a}
\ea
and
\ba
\tilde L_{ab}{}^c & = &  \ha (\tilde M_s^{-1})^{cd}\left(
\tilde B_{ab}{}^e (M^{-1})_{ed} + \tilde A_{db}{}^e (M^{-1})_{ae}- \tilde A_{ad}{}^e (M^{-1})_{eb}
  \right) \ ,
\nonumber\\
\tilde R_{ab}{}^c & = & \ha (\tilde M_s^{-1})^{cd}\left( \tilde A_{ab}{}^e (M^{-1})_{de}
+ \tilde B_{ad}{}^e (M^{-1})_{eb} - \tilde B_{db}{}^e (M^{-1})_{ae}
\right) \ ,
\label{kkll1}
\ea
where\footnote{
Note also the identities
\be
R^{ab}{}_b  = \tilde f^{ab}{}_b +  M^{ab} f_{bc}{}^c + (MM_s^{-1}M)^{bc} f_{bc}{}^a\ ,
\qq
R^{ba}{}_b  =  0\
\ee
and
\be
L^{ab}{}_b  =  0\ ,
\qq
L^{ba}{}_b  = -\tilde f^{ab}{}_b +  M^{ba} f_{bc}{}^c - (MM_s^{-1}M)^{bc} f_{bc}{}^a\ .
\ee
Similar relations hold for the tilded dual quantities.
}

\be
M_s = \ha (M+M^T) \ ,\qq \tilde M_s = \ha \left[(M^{-1}) + (M^{-1})^T\right]\ .
\ee

\no
From the results in \cite{ValCli} we deduce that the
one-loop renormalization group flow system of equations corresponding to \eqn{action1} is
\be
{d M^{ab}\ov dt}
= {\l \ov 2\pi } R^{ac}{}_d L^{db}{}_c\ .
\label{rg1}
\ee
Similarly, for its dual \eqn{action2} we have
\be
{d (M^{-1})_{ab}\ov dt}
= {\l \ov 2\pi }\tilde R_{ac}{}^d \tilde L_{db}{}^c\ ,
\label{rg2}
\ee
Also it turns out that the overall coupling $\l$ does not get renormalized, as noticed already
in a particular example in \cite{PLsfe4}.
Hence, we may, in the remaining of this paper absorb, for notational convenience,
the overall factor $\displaystyle{\l\ov 2\pi}$ into a redefinition of $t$.

\subsection{Proof of compatibility}

In order to show that \PL\ folds at the one-loop quantum level we have
to demonstrate compatibility with the renormalization flow equations.
This will be explicit if the
two systems \eqn{rg1} and \eqn{rg2} are if fact equivalent.

\no
We first note the, easy to prove, relations
\be
A^{ab}{}_c = -M^{db} \tilde A_{cd}{}^a\ ,\qq B^{ab}{}_c = M^{ad} \tilde B_{dc}{}^b \ ,
\ee
their dual (which give no further information)
\be
\tilde A_{ab}{}^c = -(M^{-1})_{db}  A^{cd}{}_a\ ,\qq \tilde B_{ab}{}^c = (M^{-1})_{ad}  B^{dc}{}_b
\label{tatt}
\ee
and that
\be
M_s = M \tilde M_s M^T = M^T \tilde M_s M \ .
\label{tatt1}
\ee

\no
Next we note the useful identities
\be
R^{ab}{}_c = M^{ad} M^{eb} (M^{-1})_{fc} \tilde R_{de}{}^f\ ,\qq
L^{ab}{}_c = -M^{ad} M^{eb} (M^{-1})_{cf} \tilde L_{de}{}^f\ .
\label{fjkk}
\ee
To prove them we start form the right hand side in the first identity
which using \eqn{tatt} and \eqn{tatt1} can be cast into the following form
\be
\ha (M_s^{-1})_{cd}\left(-M^{ae} A^{db}{}_e + M^{de} B^{ab}{}_e - M^{eb}B^{da}{}_e\right)\ .
\label{endiam}
\ee
To proceed we note that from the definitions \eqn{fhh11} we have
\ba
A^{db}{}_e & = & B^{db}{}_e - M^{df} f_{fe}{}^b - M^{fb} f_{ef}{}^d\ ,
\nonumber\\
B^{ab}{}_e & = & A^{ab}{}_e + M^{af}f_{fe}{}^b + M^{fb} f_{ef}{}^a\ ,
\\
B^{da}{}_e & = & - B^{ad}{}_e + M^{af} f_{fe}{}^d + M^{df} f_{fe}{}^a\ .
\nonumber
\ea
Substituting into \eqn{endiam} we obtain
\ba
&&\ha (M_s^{-1})_{cd}\left( A^{ab}{}_e M^{d e} + B^{ad}{}_e M^{eb} - B^{db}{}_e M^{ae}\right)
\nonumber\\
&& +  \ha (M_s^{-1})_{cd}\Big( f_{fe}{}^b M^{ae} M^{df} + f_{ef}{}^d M^{ae} M^{fb}
\\
&& \phantom{xxxxx}
+f_{fe}{}^b M^{de} M^{af} + f_{ef}{}^a
M^{de}M^{fb} - f_{fe}{}^d M^{eb}M^{af} - f_{fe}{}^a M^{eb}M^{df}\Big)\ .
\nonumber
\ea
The parenthesis in the first line is simply $R^{ab}{}_c$, whereas the remaining terms vanish.
The second identity in \eqn{fjkk} follows from a similar computation.
Alternatively, one can prove it by noticing the transformation
\be
M\to -M^T\qq \Longrightarrow \qq  R^{ab}{}_c\longleftrightarrow - L^{ba}{}_c\ ,\quad
\tilde R_{ab}{}^c\longleftrightarrow - \tilde L_{ba}{}^c \ .
\label{3.19}
\ee
Using \eqn{fjkk} one easily sees that the system \eqn{rg1} implies \eqn{rg2} and
vice versa.

\no
Hence, we conclude that, at one-loop in perturbation theory, general $\s$-models related
by Poisson--Lie T-duality are equivalent under the renormalization group flow.

\section{Properties of renormalization group flow}

In this section we study several properties of the renormalization group equations.
First note that, setting the matric $M$ proportional to the identity is not consistent with the
renormalization group equations for general Drinfeld doubles. It is, however, consistent for
abelian $\tilde G$'s,
in the basis where the identity is the invariant metric for
the group $G$.

\no
Consider next a subgroup $H$ of $G$ and another $\tilde H$, of equal dimension, subgroup
of $\tilde G$. We split the index $a=(i,\a)$,
where $i=1,2,\dots , \dim(H)$ and $\a = 1,2,\dots , \dim(G/H)$.
We note in passing that, $H$ and $\tilde H$ form a Drinfeld double as well. This
can be easily seen by restricting the four free indices in
the mixed Jacobi identity in \eqn{ftilf1} to the subgroup.

\no
We will use the following notation for the matrices $M^{ab}$ and  $\Pi^{ab}$
\be
(M^{ab}) = \left(%
\begin{array}{cc}
  H^{ij} & M^{i \b} \\
  M^{\a j} & K^{\a\b} \\
\end{array}%
\right)
\ ,\qq
(\Pi^{ab}) = \left(%
\begin{array}{cc}
  \Pi_0^{ij} & \Pi_2^{i\a} \\
  -\Pi_2^{j\b} & \Pi_1^{\a\b} \\
\end{array}%
\right)\ .
\label{kl11}
\ee
and similarly for $\tilde \Pi_{ab}$.

\subsection{Consistent truncation of the parameter space}

A natural question to investigate is to what a extend we may choose in the matrix \eqn{kl11}
the mixed-index elements to be zero, namely that
\be
M^{i\a} = M^{\a i} = 0 \ .
\label{kler3}
\ee
We will find the conditions under which the matrix $M$ remains of a block diagonal form,
under the renormalization group flow.
In general we have
\ba
 {dM^{i\a}\ov dt} = R^{ic}{}_d L^{d\a}{}_c
=
 R^{ij}{}_k L^{k\a}{}_j + R^{i\b}{}_k L^{k\a}{}_\b
+ R^{ij}{}_\b L^{\b\a}{}_j + R^{i\b}{}_\g L^{\g\a}{}_\b \ .
\label{jdoo1}
\ea
With the choice \eqn{kler3} and using that $f_{ij}{}^\a = f_{k\b}{}^l =0 $,
it is easy to compute that the quantities defined in \eqn{fhh11}-\eqn{kkll1} are zero
when two indices are latin letters and one is a greek one.
In addition, the form of $L^{\g\a}{}_\b$ is that of \eqn{kkll1a} with the latin-letter
indices replaced by greek ones, hence
projected completely into the coset space.
We conclude that, in the right hand side of \eqn{jdoo1} only the last term is non-vanishing.
Hence, a sufficient condition to preserve the choice \eqn{kler3} under renormalization group
flow is that the coset spaces $G/H$
and $\tilde G/\tilde H$ are
symmetric, i.e. the structure constants $f_{\a\b}{}^\g =\tilde f^{\a\b}{}_\g=0$. This is not
a necessary condition as well, even in the abelian $\tilde G$ case. An example is
the non-symmetric coset space $SU(3)/SU(2)$. We will present the details in section 5.

\no
Finally, for later use, we write the factor
\be
R^{i\b}{}_\g = \ha (K_s^{-1})_{\g\d}\left(H^{il}(f_{l\zeta }{}^\d K^{\zeta \b}
+ f_{l\zeta }{}^\b K^{\d \zeta } - \tilde f^{\d \b}{}_l) + K^{\d \zeta}(\tilde f^{i \b}{}_\zeta
- f_{\zeta \eta}{}^i K^{\eta \b})
-\tilde f^{i \d}{}_\zeta K^{\zeta \b}\right)\ .
\label{jjfj}
\ee

\subsection{Generalized coset spaces}

Let's examine the behaviour and equivalence under renormalization group flow
of a class of $\s$-models introduced at the
purely classical level in \cite{PLsfe4}. Consider the limit
\be
H^{ij}\to \infty\quad \Longleftrightarrow \quad (H^{-1})_{ij}\to 0\ ,
\label{eoin}
\ee
in a uniform way for all matrix elements, meaning that ratios of matrix
elements remain constant in this limit.
Then, the action \eqn{action1} takes the form
\be
S= {1\ov 2 \l} \int \S_{\a\b} L^\a_\m L^\b_\n \del_+ X^\m \del_- X^\n\ .
\qq \S = (K - \Pi_1)\inv\ .
\label{actn1}
\ee
Since
\be
M^{-1} = \left(
           \begin{array}{cc}
             0 & 0 \\
             0 & K^{-1} \\
           \end{array}
         \right) + {\cal O}\left( H^{-1}\right)\ ,
\ee
the dual action \eqn{action2} becomes
\be
\tilde S= {1\ov 2 \l} \int \tilde \S^{AB} \tL_{A\m} \tL_{B\n} \del_+
 \tilde X^\m \del_- \tilde X^\n\ ,
\qq \tilde \S =
\left(
  \begin{array}{cc}
    \tilde \Pi_0 & \tilde \Pi_2 \\
    -\tilde \Pi_2 & K^{-1} -\tilde \Pi_1 \\
  \end{array}
\right)\ .
\label{actn2}
\ee
Upon taking the limit \eqn{eoin} the number of variables in the two actions
has been reduced to $\dim(G/H)$.
The reduced dimensionality of \eqn{action1} happens since,
after taking the limit \eqn{eoin}, a local gauge invariance develops provided that
certain conditions hold.
In particular, it has been shown that \eqn{actn1} is invariant under the local gauge transformation
$g\to g h$, with $h \in H$, provided that the following conditions hold \cite{PLsfe4}
\be
\tilde f^{\a\b}{}_i = f_{i\g}{}^\a K^{\g\b} + f_{i\g}{}^\b K^{\a\g}\ .
\label{rel1}
\ee
Note that for abelian group $\tilde G$ this reduces to the usual condition in cosets $G/H$
for an invariant tensor \cite{cosetcoinv} so that \eqn{rel1} presents the analog
of this condition for our generalized cosets. Also, when $\tilde G$ is non-abelian
it is not consistent to take $K$ to be a symmetric matrix.
Then, we may gauge-fix the
$\dim(H)$ parameters in the group element $g\in G$.
The most efficient way, that completely fixes the gauge,
is to parametrize as $g=\kappa h$, where $h\in H$ and $\kappa\in G/H$, and then set $h=I$.

\subsubsection{Renormalization group flow}

We would like first to investigate if the limit \eqn{eoin} is consistent with the
renormalization group equations
\eqn{rg1}, \eqn{rg2}. In general
\ba
 {dK^{-1}_{\a\b}\ov dt} =  \tilde R_{\a c}{}^d \tilde L_{d\b}{}^c =
 \tilde R_{\a i }{}^j  \tilde L_{j\b}{}^i + \tilde R_{\a i}{}^\g  \tilde L_{\g \b}{}^i
+ \tilde R_{\a\g}{}^i \tilde L_{i\b}{}^\g  + \tilde R_{\a\g}{}^\d \tilde L_{\d\b}{}^\g \ .
\label{ll12}
\ea
In the limit \eqn{eoin}, $(M^{-1})_{i\a}$ and $(M^{-1})_{\a i}$ as well as
the quantities \eqn{fhh11}-\eqn{kkll1}
when two of the indices are latin letters and one is a greek one, are of order $(H^{-1})_{ij}$.
Hence, the first term in \eqn{ll12} vanishes.
Among the remaining terms the last one is actually independent of the matrix $H$.
In the second term the factor $\tilde R_{\a i}{}^\g$ has a finite part and a vanishing part under
the limit \eqn{eoin}.
Similarly, the second factor $\tilde L_{\g \b}{}^i$ has a divergent part and a
finite part under this limit. The divergent part is
\be
\ha (\tilde H_s^{-1})^{ij}\left[(f_{j\b}{}^\d - \tilde f^{\d \e}{}_j K^{-1}_{\e\b}) K^{-1}_{\g\d}
- f_{\g j}{}^\d K^{-1}_{\d\b}\right]\ .
\ee
It can be easily seen that it vanishes upon using \eqn{rel1}.
Therefore, to compute in the limit \eqn{eoin}, the term $\tilde R_{\a i}{}^\g \tilde L_{\ g \b}{}^i$
we need to keep
the finite part of both factors. With some rearrangement we obtain
\be
{1\ov 4} (H  H^{-1}_s)^i{}_k \left[ f_{\a i}{}^\e K^{-1}_{\d \e} - K^{-1}_{\a\e}
(f_{\d i}{}^\e + K^{-1}_{\d \g} \tilde f^{\g \e}{}_i)\right ]
\left[ f_{\g\b}{}^k + K^{-1}_{\g\zeta} \tilde f^{\zeta k}{}_\b + K^{-1}_{\zeta\b} \tilde f^{\zeta k}{}_\g
\right]\ .
\ee
Using \eqn{rel1} we finally obtain
\be
(H  H^{-1}_s)^i{}_k \G_{ i\a\b }{}^k \ ,
\ee
where
\be
\G_{ i\a\b }{}^k =
\ha f_{i\eta}{}^\d K^{\eta \g} (K^{-1})_{\a\d}
( f_{\g\b}{}^k + K^{-1}_{\g\zeta} \tilde f^{\zeta k}{}_\b + K^{-1}_{\zeta\b} \tilde f^{\zeta k}{}_\g)\ .
\ee
To this term we should add a similar one coming from the third term in \eqn{ll12}
which using the transformation \eqn{3.19} reads
\be
(H^T  H^{-1}_s)^i{}_k \D_{ i\a\b }{}^k \ ,
\ee
where
\be
\D_{ i\a\b }{}^k =
\ha f_{i\eta}{}^\d K^{\g\eta}(K^{-1})_{\d \b}
( f_{\g\a}{}^k - K^{-1}_{\zeta \g} \tilde f^{\zeta k}{}_\a - K^{-1}_{\a\zeta} \tilde f^{\zeta k}{}_\g)\ .
\ee
It is clear that, in order for the limit \eqn{eoin} to be independent of the form of the matrix $H$
the latter has to be
symmetric.\footnote{Alternatively, we may slightly modify the limit \eqn{eoin} to involve
only the symmetric part of $H$.
}
Denoting by
\be
 \G_{\a\b}= \G_{i\a\b}{}^i \ ,\qq \D_{\a\b}=  \D_{i\a\b}{}^i\ ,
\ee
we finally have that, in the limit \eqn{eoin}
\ba
 {dK^{-1}_{\a\b}\ov dt}=
 \G_{\a\b} +  \D_{\a\b} +
\tilde R_{\a\g}{}^\d \tilde L_{\d\b}{}^\g  \ .
\label{kl1}
\ea
This would have been the result for the renormalization group flow system of equations
for the dual coset model \eqn{actn2}. Note that for symmetric spaces the last term
is zero and
the running is solely due to the first two terms. These could be thought
of as a remnant of the
original full group structure in \eqn{action2}.
Had we performed a similar limiting procedure starting from the system \eqn{rg1}
corresponding to the
action \eqn{actn1} we would have obtained
\ba
 {dK^{\a\b}\ov dt}=
 P^{\a\b} + Q^{\a\b} +
 R^{\a\g}{}_\d L^{\d\b}{}_\g  \ ,
\label{crg1}
\ea
where
\be
 P^{\a\b}= P_i{}^{\a\b i} \ ,\qq Q^{\a\b}=  Q_i{}^{\a\b i}\ ,
\ee
with
\ba
P_i{}^{\a\b k} &  = & \ha f_{i\g}{}^\a
\left[K^{\g\eta}(\tilde f^{k\b}{}_\eta - f_{\eta \zeta}{}^k K^{\zeta \b}) +\tilde f^{k \g}{}_\zeta
K^{\zeta \b} \right]\ ,
\nonumber\\
Q_i{}^{\a\b k} &  = &- \ha f_{i\g}{}^\b
\left[K^{\eta\g}(\tilde f^{k\a}{}_\eta + f_{\eta \zeta}{}^k K^{\a\zeta }) +\tilde f^{k \g}{}_\zeta
K^{\a\zeta } \right]\ .
\label{crg2}
\ea

\subsubsection{One loop equivalence}

The equivalence of the systems \eqn{kl1} and \eqn{crg1} can be demonstrated by noting that
there exist expressions similar to \eqn{fjkk} projected completely to the coset space indices,
namely that
\be
R^{\a\b}{}_\g = K^{\a\d} K^{\eta \b} (K^{-1})_{\zeta \g} \tilde R_{\d \eta}{}^\zeta\ ,\qq
L^{\a\b}{}_\g = -K^{\a\d} K^{\eta \b} (K^{-1})_{\eta \zeta} \tilde L_{\d\eta}{}^\zeta\
\label{fjkkco}
\ee
and in addition one may prove that
\be
P^{\a\b} = - K^{\a\g} K^{\d\b} \G_{\g\d}\ ,\qq Q^{\a\b} = - K^{\a\g} K^{\d\b} \D_{\g\d}\ .
\ee

\no
Next we investigate when the condition \eqn{rel1} for being able to take the limit \eqn{eoin}
consistently, is preserved under the renormalization group flow. We demand that
\be
f_{i\g}{}^\a {d K^{\g\b} \ov dt} + f_{i\g}{}^\b {d K^{\a\g}\ov dt} = 0 \ .
\label{ftkf}
\ee
The only possibility that this is obeyed is that the right hand side of \eqn{crg1} is an
invariant tensor obeying a condition similar to \eqn{rel1}.
In the case with $\tilde{f}^{ab}{}_c=0$, \eqn{ftkf}
can be proven by repeatedly using the Jacobi identity for the $f_{ab}{}^c$'s, the coset
constraint on $K^{\a\b}$ \eqn{rel1} and by taking $K_s^{\a\b}$ to
be block diagonal. This can always be done by a proper transformation of the group element $g\in G$ on the
action.\footnote{In the proof we also use the fact that
\be
f_{i\g}{}^\eta f_{c\eta}{}^d f_{d\zeta}{}^c\ ,
\qq f_{i\g}{}^\varepsilon f_{\d\varepsilon}{}^\eta f_{\eta\zeta}{}^\d\ ,
\ee
are antisymmetric under the interchange of $\g$ and $\zeta $.}
For the case with $\tilde f^{ab}{}_c\neq 0$ the proof is much more complicated and we have just explicitly
checked that this is indeed true for the cosets of subsection 5.1 below.

\no
Finally, we note that the limit \eqn{eoin} is not a fixed point of the renormalization group flow in general.
Consider the special case of $H^{ij}= H \d^{ij}$ with $\d^{ij}$ being the invariant metric of the group.
Then after some computations we obtain that
\be
{d H^{-1}_{ij}\ov dt } = {1\ov 4} f_{i kl}f^{jkl} - {1\ov 4}  (K^{-1}_s)_{\a\g} (K^{-1}_s)_{\b\d}
\tilde f^{\a\b}{}_i \tilde f^{\g\d}{}_j + {\cal O}\left( H^{-1}\right)\ .
\label{hinve}
\ee
Note that the matrix remains under renormalization group flow symmetric, but no longer diagonal.
Also it is zero for abelian subgroups $H$, when simultaneously $\tilde f^{\a\b}{}_i =0$.

\no
The off diagonal elements do also flow even in the coset limit unless the space is
symmetric. Note that the flow is well defined since the coefficient of the $H$-dependent term
in \eqn{jjfj} vanishes thanks to the condition \eqn{rel1}.

\section{Examples}

\subsection{Six-dimensional doubles}

In this section we first construct several examples
based on a six-dim Drinfeld double decomposition, into two
three-dimensional Lie algebras.
The associated three-dimensional groups, $G$ and $\tilde{G}$ have generators denoted
by ${T_a}$ and ${\tilde{T}^a}$, where $a=1,2,3.$
It is convenient to split the index $a=(3,\a)$, with $\a=1,2$.
Abelian subgroups are generated by $T_3$ and $\tilde T^3$, so that $\a$
takes values in the corresponding two-dimensional coset spaces.
The non-vanishing structure constants of the algebras next to be considered are
\ba
  IX:&& f_{\a\b}{}^3=f_{3\a}{}^\b=\e_{\a\b}\ ,\qq
V:\quad  f_{3\a}{}^\b=\d_{\a\b}\ ,
\nonumber\\
  VII_0: && f_{3\a}{}^\b=\e_{\a\b}\ ,
\phantom{xxxxxxxx}\
II: \quad f_{\a\b}{}^3=\e_{\a\b}\ ,
\ea
where we have labeled them according to the standard Bianchi classification for three
dimensional algebras.
Only four combinations of the above correspond to six-dim Drinfeld doubles, namely
$(IX,V),(VII_0,II),(II,V),(V,VII_0)$, in a $(G,\tilde G)$ notation.
The renormalization group equations for the first pair has been constructed in
\cite{PLsfe3} whereas a classification of all six-dimensional doubles based
on the Bianchi classification of three-dimensional algebras can be found in
\cite{Jafarizadeh} (see also \cite{Snobl} for further related issues).

\no
Since $f_{\a\b}{}^\g=\tilde f^{\a\b}{}_\g=0$, we may according to the
results of subsection 4.1 take consistently
the form of the matrix $M^{ab}$ as block diagonal, namely
\ba
(M^{ab})=\left(
  \begin{array}{cc}
    1/G & 0 \\
    0 & K^{\a\b} \\
  \end{array}
\right)
\ , \qq
(K^{\a\b})=
\left(
  \begin{array}{cc}
    A & B \\
    C & D \\
  \end{array}
\right)\ .
\label{kl22}
\ea
Taking the coset limit \eqn{eoin}, whenever possible, corresponds to $G\to 0$.

\no
\underline{The case of $(IX,V)$:}
This case was examined in detail in \cite{PLsfe4}.
The beta-function equations for the general matrix in \eqn{kl22}, are quiet lengthy
and they will not be presented here.
However, for $K^{\a\b}$ satisfying \eqn{rel1}, the expressions become much simpler.
Setting $A=D=a$, $B=-C=b-1$ and $G=(1+g)/a$ we find
\ba
&&{da\ov dt}={1+a^2-b^2\ov 2a^2}((g-1)a^2+(g+1)(b^2-1))\ ,
\nonumber \\
&&{db\ov dt}={b\ov a}(a^2(g-1)+(g+1)(b^2-1))\ ,
\label{ixv}\\
&&{dg\ov dt}={1+g \ov a}(g(1+a^2)+(g+2)b^2)\ ,
\nonumber
\ea
which are precisely the expressions in eq. (4.4) in \cite{PLsfe4}. The coset limit
is $g\to -1$.

\no
\underline{The case of $(VII_0,II)$:}
It turns out that in this case the coset construction is not possible, i.e. \eqn{rel1} has no solution.
To simplify the renormalization group flow,
we present just a consistent truncation of the beta-function equations namely, along $A=D$ and
$B=-C$
\be
{dA\ov dt}=-{1\ov 2 A G}\ , \quad {dB\ov dt}=0 \ , \quad {dG\ov dt}= -{1\ov 2 A^2}\ .
\ee

\no
\underline{The case of $(II,V)$:}
In this case there is no constraint on the coset construction,
i.e. no constraint on $K^{\a\b}$.
We present just a consistent truncation of these equations namely, along $A=D,B=-C$
\ba
&&{dA\ov dt}={(A^4 - B^2 (2 + B)^2)\ov 2 A}G \ ,
\nonumber \\
&&{dB\ov dt}=(1 + B) (A^2 + B (2 + B)) G\ ,
\\
&&{dG\ov dt}= {(A^2 + B^2) (A^2 + (2 + B)^2)\ov 2 A^2} G^2\  .
\nonumber
\ea

\no
\underline{The case of $(V,VII_0)$:}
A consistent truncation of the beta-function equations is along $A=D$ and $B=-C$
\be
{dA\ov dt}=-{2 B(B + A G)\ov A G}\ ,\quad
{dB\ov dt}=2 \left(A + {B\ov G}\right)\ ,\quad
{dG\ov dt}= -2 - {2 B^2\ov A^2}\  .
\ee

\subsection{Flow in $SU(3)$ and the coset $SU(3)/SU(2)$}

We use the structure constants in the Gell--Mann basis for $SU(3)$ (see, for instance, eq. (5.2) of the second of
\cite{cosetcoinv})
\be
f_{12}{}^3=2\ ,\quad f_{14}{}^7=-f_{15}{}^6=f_{24}{}^6=f_{25}{}^7=f_{34}{}^5=-f_{36}{}^7=1\ ,\quad f_{45}{}^8=f_{67}{}^8=\sqrt{3}\ ,
\ee
where the rest are obtained by antisymmetrization
and pick up as an $SU(2)$ subgroup the one generated by $T_i$, $i=1,2,3$.
Then in is easy to check that the most general invariant matrix $K$ is
\be
K = \left(
  \begin{array}{ccccc}
    A & C & D & Z & 0 \\
    -C & A & Z & -D & 0 \\
    -D & -Z & A & C & 0 \\
    -Z & D & -C & A & 0 \\
    0 & 0 & 0 & 0 & B \\
  \end{array}
\right)\ ,
\ee
which has diagonal symmetric part as well as an antisymmetric one.

\no
For the renormalization group flow in the full $SU(3)$ model we take for the matrix $M$ the form
\eqn{kl11} with $M^{i\a}=M^{\a i}=0$ and $H= {1\ov G}\ \mathbb{Id}_{3\times 3}$. We find that the system
\eqn{rg1} leads to a consistent system and in particular the r.h.s. of \eqn{kler3}
is zero even though the coset $SU(3)/SU(2)$ is not a symmetric space. We present the equations in the
particular case of zero antisymmetric part, i.e. when $C=D=Z=0$, which is a consistent truncation. We obtain
\be
{dA\ov dt}= {3 A^2\ov 2} {A(B G+1)-4 B\ov B }\ ,\qq {dB\ov dt} = - 3 A^2\ ,\qq {dG\ov dt} = G^2 A^2 +2\ .
\ee
For the coset $SU(3)/SU(2)$ the flow equations are the first two in the limit $G\to 0$.

\no
Similar results can be found for the models $SO(4)/SO(3)$ and $Sp(4)/(SU(2)\times U(1))$ cases as well,
but we will not present the details.

\section{Future directions}

It would be very interesting to formulate the renormalization group
flow in a duality-invariant way. For this an appropriate starting point should be
the duality-invariant action formulation of \PL\ in \cite{DriDou}. Since this action has twice as
many fields as \eqn{action1} and \eqn{action2} and in addition in lacks two-dimensional Lorentz
invariance, the corresponding one-loop counter-terms should be derived as a necessary first step.

\no
In trying to explicitly solve the system of the beta-functions for low dimensional models
it helps to know the renormalization group invariants. This can be possibly worked out example by
example, for instance for the flows of the previous section.  For the system \eqn{ixv} such an
invariant was found in \cite{PLsfe4} and indeed helped in reducing it into an single,
first order non-linear differential equation.
However, for general considerations it would be interesting to
have the forms of some if not of all of such invariants and classify them using the underlying
group theoretic structure of \PL.

\no
As indicated above, in all known examples canonically equivalent classical $\s$-models
are also equivalent under one-loop renormalization group flow in the sense already explained. It is
interesting to search and provide a general proof of that statement.

\vskip  .7 cm

\centerline{\bf Acknowledgments}

\no
K. Siampos acknowledges support by the  Greek State Scholarship Foundation (IKY).


\end{document}